\documentclass[aps,prl,twocolumn,superscriptaddress]{revtex4}

\usepackage{graphicx}
\usepackage{dcolumn}
\usepackage{bm}
\def\rnum#1{\expandafter{\romannumeral #1}}
\def\Rnum#1{\uppercase\expandafter{\romannumeral #1}}

\begin{document}


\title{ Electron Paramagnetic Resonance of Boron Acceptors in Isotopically Purified Silicon}

\author{H.~Tezuka}
\affiliation{School of Fundamental Science and Technology, Keio University, Yokohama 223-8522, Japan}

\author{A.~R.~Stegner}
\affiliation{Walter Schottky Institut, Technische Universit\"{a}t
M\"{u}nchen, Am Coulombwall 3, 85748 Garching, Germany}

\author{A.~M.~Tyryshkin}
\affiliation{Department of Electrical Engineering, Princeton University, Princeton, NJ 08544, USA}

\author{S.~Shankar}
\affiliation{Department of Electrical Engineering, Princeton University, Princeton, NJ 08544, USA}

\author{M.~L.~W.~Thewalt}
\affiliation{Department of Physics, Simon Fraser University, Burnaby, British Columbia, Canada V5A 1S6}

\author{S.~A.~Lyon}
\affiliation{Department of Electrical Engineering, Princeton University, Princeton, NJ 08544, USA}

\author{K.~M.~Itoh}
\email[e-mail: ]{kitoh@appi.keio.ac.jp}
\affiliation{School of Fundamental Science and Technology, Keio University, Yokohama 223-8522, Japan}

\author{M.~S.~Brandt}
\affiliation{Walter Schottky Institut, Technische Universit\"{a}t
M\"{u}nchen, Am Coulombwall 3, 85748 Garching, Germany}
\date{\today}

\begin{abstract}
The electron paramagnetic resonance (EPR) linewidths of B
acceptors in Si are found to reduce dramatically in
isotopically purified $^{28}$Si single crystals. Moreover,
extremely narrow substructures in the EPR spectra are visible
corresponding to either an enhancement or a reduction of the
absorbed microwave on resonance. The origin of the substructures
is attributed to a combination of simultaneous double excitation
and spin relaxation in the four level spin system of the
acceptors. A spin population model is developed which
qualitatively describes the experimental results.
\end{abstract}


\maketitle

\begin{figure}[b]
\includegraphics[width=8.5cm]{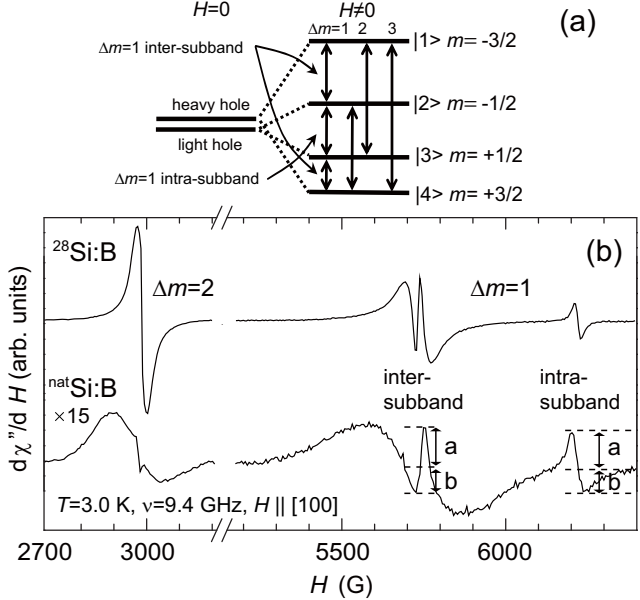}
\caption{\label{figure1} (a) The level schemes of B acceptors. An
externally applied magnetic field splits the heavy (light) hole
subband into two energy states $|\pm 3/2\rangle$ ($|\pm
1/2\rangle$).  We refer to each level as $|1\rangle$-$|4\rangle$
in the order shown.  On top of the EPR-allowed transitions defined
by $\Delta m$=1, the experimentally observed nominally
EPR-forbidden transitions $\Delta m$=2 and $\Delta m$=3 are also
shown. (b) EPR spectra recorded at $T$=3.0~K on $^{28}$Si:B and
$^{\mathrm{nat}}$Si:B samples showing the $\Delta m$=1 and 2
transitions. }
\end{figure}

Boron is the most widely employed shallow acceptor in silicon.
However, due to the existence of four degenerate energy levels
originating from the light and heavy hole bands (shown in
Fig.~\ref{figure1}~(a)) and their strong sensitivity to
perturbations like internal strain~\cite{Kohn}, B exhibits
rich optical and electron paramagnetic resonance (EPR) spectra,
which are still not fully understood. Recently, some open
questions associated with the optical properties of this acceptor
 have been solved due to the availability of isotopically enriched
Si~\cite{SiIsotope1, SiIsotope2}. Using such samples, 
it has been shown that the random distribution of
the different host Si isotopes causes a residual acceptor
ground state splitting in natural Si which is experimentally
observed in the photoluminescence (PL) spectra of acceptor bound
excitons~\cite{KaraPL} and also gives rise to inhomogeneous
broadening of the infrared absorption
spectra~\cite{KaraIR,KaraSim}.

Similarly, host isotope effects might also account for various
properties of the EPR spectra of B in Si that have not been
understood so far. The first successful boron EPR required the
application of large homogeneous external stress to diminish the
influence of inhomogeneous internal strain fields that are mainly
induced by randomly distributed point defects and dislocations.
These defects lead to a statistical distribution of ground state
splittings, which, depending on the defect concentration, can
result in an extremely large inhomogeneous broadening of the EPR
resonance lines in externally unstrained Si~\cite{Feher}. The
first observation of boron EPR without external stress was finally
realized in 1978~\cite{Neubrand1, Neubrand2}, when Si crystal
growth techniques had developed enough to obtain samples of
sufficient crystalline quality. However, the experimentally
observed EPR linewidths have always been much broader
than could be explained solely by the residual defect
concentrations in high quality Si. In addition, Neubrand observed
sharp substructures in the main resonance peaks, with and without
externally applied stress~\cite{Neubrand1, Neubrand2}. More recent
studies on acceptor EPR probing dependencies on temperature,
external stress, and frequency~\cite{Kopf, Scad} found the same
sharp features, but the origin of these substructures remained
unclear.

Here, we present experimental evidence that indeed the random
distribution of stable Si isotopes is the major cause of
inhomogeneous broadening of boron EPR spectra in natural Si.
Moreover, thanks to the sharpening of the main EPR lines, the
improved visibility of the sharp substructures allows for detailed
studies of their origin.  The substructures are shown to arise
from an interplay between simultaneous excitation of two EPR
transitions and spin relaxation in the four level spin system of B
acceptors in Si.  A theoretical model based on this assumption is
developed to describe the temperature dependence of the
substructures.

The experiments were performed with two $p$-type Si single
crystals. The first sample, referred to as $^{\mathrm{nat}}$Si:B,
was float-zone grown, has a natural isotopic composition and a
B concentration of 1.4$\times$10$^{14}$~cm$^{-3}$. The
isotopically purified sample ($\simeq$99.98~\% $^{28}$Si),
referred to as $^{28}$Si:B, comes from the neck of a float-zone
crystal and is doped with B to a concentration of 3($\pm$1)$\times$10$^{14}$~cm$^{-3}$.

EPR measurements at a temperature of $T$=1.8~K were performed
using a Bruker E580 EPR spectrometer equipped with a flex-line resonator
(ED-4118MD5) operating at a microwave frequency $\nu$=9.6~GHz. 
An Oxford CF935 helium-flow cryostat was used to
maintain the temperature. For experiments at $T$$\geq$2.8~K, a
Bruker E500 spectrometer in conjunction with a super-high-Q
resonator (ER-4122SHQE) was used at $\nu$=9.4~GHz. In this case,
the sample was cooled using an Oxford ESR900 helium-flow cryostat.
Lock-in detection using magnetic field modulation leads to the
detection of the first derivative $d\chi'' / dH$ of the imaginary
part of the magnetic susceptibility $\chi''$ with respect to the
magnetic field $H$. The absorption curves, which are proportional
to $\chi''$, are obtained via numerical integration of the
measured spectra.

Figure~\ref{figure1}~(b) shows two overview EPR spectra that were
measured on the $^\mathrm{28}$Si:B and the $^\mathrm{nat}$Si:B
samples. In both spectra, we observe three distinct B-related
resonances. From the angular dependence of the effective
g-values~\cite{Neubrand1,Neubrand2}, we assign the resonance at
$H$=2980~G to a superposition of the two inter-subband $\Delta
m$=2 transitions, where $m$ is the magnetic quantum number of the
total angular momentum $J$=3/2 of the B acceptor. The
resonances at $H$=5730~G and $H$=6220~G originate from a
superposition of the two inter-subband $\Delta m$=1 transitions
$|1\rangle$$\leftrightarrow$$|2\rangle$ and
$|3\rangle$$\leftrightarrow$$|4\rangle$ and the intra-subband
$\Delta m$=1 transition $|2\rangle$$\leftrightarrow$$|3\rangle$,
respectively~\cite{Neubrand1, Neubrand2, Luttinger}. The most
prominent difference between the two spectra is the decreased
linewidth of all B-related resonances in the isotopically purified
sample. This effect is most strongly pronounced for inter-subband
transitions. As found by PL in Ref.~\cite{KaraPL,KaraSim}, local
fluctuations of the valence band edge due to the random
distribution of the different Si isotopes are responsible for the
residual splitting of the B acceptor ground state in
$^\mathrm{nat}$Si in absence of an external magnetic field. A
detailed modelling of the influence of the isotope-induced
fluctuations on the EPR resonances shows that this effect also
accounts quantitatively for the inhomogeneous broadening of
B-related EPR signals that could not be explained by point
defect-induced random strain fields. This is in marked difference
to the broadening of donor EPR lines in Si, which is caused by
superhyperfine interaction with $^{29}$Si nuclei ~\cite{Feher2}.
The quantitative investigation of the broadening mechanism in Si:B
is beyond the scope of this paper and will be given elsewhere.
Here, we focus on the origin of the sharp substructures that
appear in the middle of the inter-subband resonances. 
The observation of these substructures in
$^\mathrm{28}$Si:B establishes that they do not result from an
isotopic effect.

\begin{figure}[b]
\includegraphics[width=8.5cm]{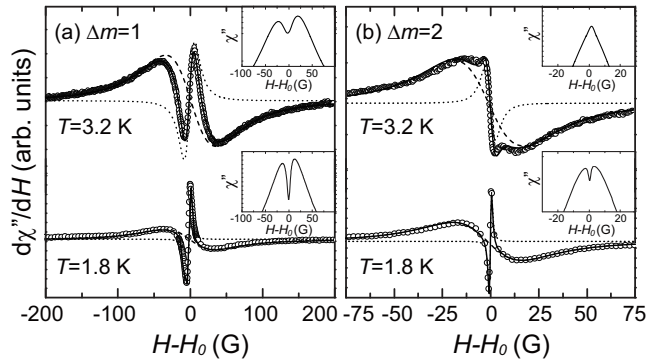}
\caption{\label{figure2} EPR spectra of the inter-subband $\Delta
m$=1 resonance (a) and the $\Delta m$=2 resonance (b) in
$^{28}$Si:B for $H||$[100]. The upper spectrum in each figure was
measured at $T$=3.2~K, $\nu$=9.38~GHz and a microwave power of
$P$=0.2~mW. The lower spectra were measured at $T$=1.8K,
$\nu$=9.56~GHz and $P$=0.002~mW. The values of $P$ were chosen to achieve the same microwave excitation rate $w$ in the two resonators used.
Solid lines show the results of numerical fittings composed of two Lorentzian lines, which are
separately shown as dashed and dotted lines. $H_0$ is the
resonance field of each signal. The amplitudes of the spectra were
normalized. The insets show the integrated absorption spectra in
the magnetic field region of the substructures in arbitrary units.}
\end{figure}

Figure~\ref{figure2} shows the inter-subband resonances 
on an expanded magnetic field scale as
measured for $^{28}$Si:B at two different temperatures. The narrow
substructure is observed in all four spectra. By cooling from 3.2
to 1.8~K, the linewidth of the substructures decreases
substantially, from 15 to 3~G for $\Delta m$=1 and from 5 to 1~G
for $\Delta m$=2. In contrast, the broad underlying signals show
much weaker changes in linewidth with temperature (from 78 to 70~G
for $\Delta m$=1, 36 to 35~G for $\Delta m$=2). Thus, even though
the line shape of the broad resonances is Lorentzian, the lack of
any significant temperature dependence suggests that the lines are
inhomogeneously broadened. This broadening is attributed to the distribution of transition 
energies between different energy levels induced by random local 
strains, which are well accounted for by the $\sim$1$\times$10$^{16}$~cm$^{-3}$ 
concentration of C impurities as measured by infrared absorption 
spectroscopy on the $\sim$607~cm$^{-1}$ C local vibrational mode~\cite{Neubrand2,a}.

The insets in Fig.~\ref{figure2} show the integrated absorption
curves in the magnetic field region of the substructure. We
observe the following experimental facts: (1) For the $\Delta m$=1
spectra (Fig.~\ref{figure2}~(a)), the substructure has a negative
sign corresponding to a decrease of the microwave absorption at
the center of the broad resonance. The intensity of this dip
increases with decreasing temperature. We have exclusively
observed a negative sign of the substructure in the $\Delta m$=1
resonance for all temperatures investigated. (2) In contrast, for
the $\Delta m$=2 resonance, the microwave absorption in the
substructure is enhanced at high temperatures, giving rise to a
sharpened peak of the integrated spectrum measured at $T$=3.2~K.
However, when the temperature is decreased to 1.8~K, the sign of
the substructure line reverses, and a dip in the absorption
spectrum is observed. (3) Microwave power saturation experiments
(not shown) reveal that the saturation behavior of the
substructure lines strongly deviates from that of 
the underlying broad signals; the relative intensity of the
substructure decreases with increasing microwave power. No power
broadening is observed for the substructure lines. (4) As has been
pointed out in Ref.~\cite{Neubrand1}, the resonance fields of the
substructure lines do not change when a small external stress is
applied. (5) We find that the linewidths of the intra-subband
$\Delta m$=1 resonance and the substructure of the inter-subband
$\Delta m$=1 signal are very similar, in particular for
temperatures below 3~K. Moreover, in the $^\mathrm{nat}$Si sample,
for each orientation of $H$ (not shown) these two resonances have
the same asymmetry of the line shape, defined as the a/b ratio
in Fig.~1(b), indicating that transitions between $|2\rangle$ and $|3\rangle$ are involved at least in the generation of the $\Delta m$=1 substructure.

The experimental observations summarized above strongly argue
against a model assuming that the narrow substructure lines arise
from a partial overlap of two asymmetric lineshapes, e.g. the
$|1\rangle$$\leftrightarrow$$|2\rangle$ and
$|3\rangle$$\leftrightarrow$$|4\rangle$ transitions in case of the
$\Delta m$=1 signal, or the
$|1\rangle$$\leftrightarrow$$|3\rangle$ and
$|2\rangle$$\leftrightarrow$$|4\rangle$ transitions in case of the
$\Delta m$=2 signal, as has been already speculated in
Ref.~\cite{Neubrand1}. On the other hand, a similar
``substructure" line in the middle of a broad resonance has been
reported for Ni$^{2+}$ centers in MgO~\cite{Smith}. However, the cross-relaxation model invoked there can not account for the positive sign of the $\Delta m$=2 substructure observed here for $T$$\geq$3.2~K. In what follows, we demonstrate which interplay of spin excitation and relaxation in the four level system of Si:B causes the narrow substructure lines observed.

\begin{figure}
\includegraphics[width=8.5cm]{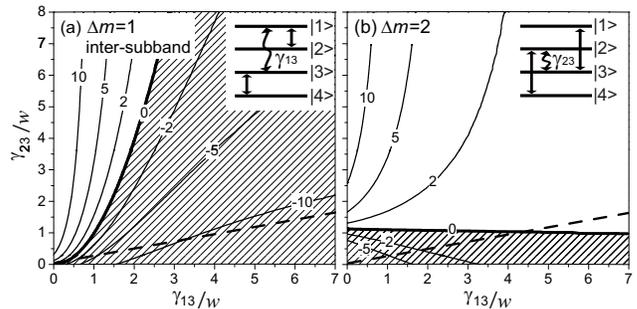}
\caption{\label{figure3}Contour plots obtained by the population
model described in the text for (a) $\Delta m$=1 and (b) $\Delta
m$=2. The contour lines indicate a relative intensity (in \%) of
the substructure lines as compared to the intensity of the broad
underlying transition. Negative and positive intensity corresponds
to a suppression (dip) and an enhancement (peak) of the microwave
power absorptions, respectively. The negative region is shown
shaded. The parameters used in the calculation were $T$=3~K,
$\nu$=9.4~GHz, and $\gamma_{14}/w$=$\gamma_{12}/w$=$1$, with $w$
being the resonant microwave excitation rate. The dashed line
represents the assumed correlated temperature dependence of
$\gamma_{13}$ and $\gamma_{23}$ in $^{28}$Si:B as discussed in the
text.}
\end{figure}

The centers of the broad resonances, where the substructure lines
develop, correspond to B acceptors that experience a strain
close to zero. In this case, the energy level scheme of the four
level system is symmetric, as shown in the insets of
Fig.~\ref{figure3}, and the energies of the transitions
$|1\rangle$$\leftrightarrow$$|2\rangle$ and
$|3\rangle$$\leftrightarrow$$|4\rangle$
[$|1\rangle$$\leftrightarrow$$|3\rangle$ and
$|2\rangle$$\leftrightarrow$$|4\rangle$] are equal, so that they
are resonantly excited at the same time. This double excitation
can lead to a change in the microwave absorption with respect to
the situation when only one transition can be excited. The
magnitude and sign of this change will depend on the ratios
between relaxation and excitation rates in the level system.

To explore this scenario, we have performed a
first calculation based on an incoherent spin population
model~\cite{Abragam, Shikata} composed of a set of rate equations.
The time evolution of the spin population $n_i$ of
each state $|i\rangle$ is given by the sum of all possible excitation and
relaxation terms. 
Relaxation and excitation rates from $|i\rangle$
to $|j\rangle$ are represented by $\gamma_{ij}$ and $w_{ij}$,
respectively.

We use our model to compare the microwave absorption between
exactly \emph{on} and slightly \emph{off resonance} conditions.
For the \emph{on resonance} condition where the substructure
emerges, $w_{12}$ and $w_{34}$ [$w_{13}$ and $w_{24}$] are assumed
to have the same non-zero value $w$ for the $\Delta m$=1 [$\Delta
m$=2] resonance, while all the other elements in the excitation
matrix are set to zero. The rate equations are solved for the 
steady state condition ($d\mathbf{n}/dt$=$0$) using the following
assumptions: (i) Boltzmann balance of the relaxation rates,
i.e.~$\gamma_{ji}$=$\gamma_{ij}\exp{(-\Delta E_{ij}/kT)}$, where
$\Delta E_{ij}$ is the energy difference between the states
$|i\rangle$ and $|j\rangle$ and $k$ is the Boltzmann constant.
(ii) $\gamma_{12}$=$\gamma_{34}$ and $\gamma_{13}$=$\gamma_{24}$,
dictated by the symmetry of the Si:B Hamiltonian when the local
strain is close to zero, (iii) $w_{ji}$=$w_{ij}$, required for
magnetic dipole transitions, and (iv) $\sum_i {n_i}$=1 to conserve
the total number of spins. The sum of the resulting population
differences, ($n_1$-$n_2$)+($n_3$-$n_4$)
[($n_1$-$n_3$)+($n_2$-$n_4$)] for the $\Delta m$=1 [for $\Delta
m$=2] transition, is used as a measure for the microwave
absorption.
To simulate the situation slightly \emph{off resonance}, where
double excitation is not possible, only one excitation at a time
is allowed, e.g.~$w_{12}$$\neq$0 [$w_{13}$$\neq$0] is assumed for
the $\Delta m$=1 [$\Delta m$=2] resonance, and the microwave
absorption represented by $n_1$-$n_2$ [$n_1$-$n_3$] is added to
the microwave absorption $n_3$-$n_4$ [$n_2$-$n_4$] that is
obtained by separately considering only the second transition with
the excitation $w_{34}$$\neq$0 [$w_{24}$$\neq$0].

Results of such calculations are shown in contour plots in
Fig.~\ref{figure3}. The relative change of the microwave
absorption due to double excitation, calculated as the population
differences between exactly \emph{on} and slightly \emph{off
resonance} conditions, is plotted as a function of the normalized
relaxation rates $\gamma_{13}/w$ and $\gamma_{23}/w$ (all other
relaxation rates were kept constant, e.g.
$\gamma_{12}/w$=$\gamma_{14}/w$=1). This calculated change in
absorption can be compared with the intensity of the substructure
in Fig.~\ref{figure2}. The open and shaded regions in
Fig.~\ref{figure3} correspond to a suppression (dip substructure)
and an enhancement (peak substructure) of microwave absorption,
respectively. The dashed line shown in the plot corresponds to a
constant ratio of $\gamma_{13} / \gamma_{23}$=4/1, motivated by
the temperature independent ratio of substructure linewidths,
$\Delta H_{\Delta m=1}/\Delta H_{\Delta m=2}$$\sim$4, observed
experimentally. Assuming all other parameters to only depend
moderately on temperature, a variation of the temperature in our
experiments can be viewed in a first approximation as a correlated
change of $\gamma_{13}$ and $\gamma_{23}$ along the dashed line.
This picture qualitatively explains the experimentally observed
temperature dependence of the substructure line intensity in
Fig.~\ref{figure2}. For example, we find a negative sign
(dip) for the $\Delta m$=1 substructure and a positive sign (peak)
for the $\Delta m$=2 substructure when $\gamma_{13}/w$$\geq $4 and
$\gamma_{23}/w$$\geq$1. Experimentally, this situation corresponds
to temperatures $\geq$3.2~K. Upon decreasing both $\gamma_{13}/w$
and $\gamma_{23}/w$ along the dashed line, corresponding to
lowering the temperature from 3.2~K to 1.8~K in the experiment, we
observe a change of the sign for the $\Delta m$=2 substructure
from positive to negative. For the $\Delta m$=1 substructure, the
sign always stays negative for all $\gamma_{13}$ and $\gamma_{23}$ along the dashed curve, agreeing with the experimental finding that the sign is
always negative at all accessible temperatures.

By simulating the model using a range of $\gamma_{ij}$ parameters
we have identified several requirements for the relative magnitude
of the $\gamma_{ij}$ in order to reproduce the experimental
observations. First, to achieve a substantial effect (e.g.~10~\%)
from the double excitation requires that all $\gamma_{ij}$ are of the
same order of magnitude and also that the applied microwave power
is close to saturation, i.e.
$\gamma_{12}$$\sim$$\gamma_{13}$$\sim$$\gamma_{14}$$\sim$$\gamma_{23}$$\sim$$w$,
as used in Fig.~\ref{figure3}. Second, to observe a temperature
dependent change of the sign for the $\Delta m$=2 substructure
requires that $\gamma_{13}$$\geq $$\gamma_{12}$. This order of the
$\gamma_{ij}$ is unusual and contrasts with
$\gamma_{12}$$\geq$$\gamma_{23}$$\gg$$\gamma_{13}$$\gg$$\gamma_{14}$
usually seen in other high-spin systems where zero-field splitting
(ZFS) dominates spin relaxation~\cite{ZFS}. However, we point out
that the ZFS Hamiltonian in Si:B is represented by the $S_i^3$
terms~\cite{Neubrand1} (in contrast to usual $S_i^2$), and
therefore the order of relaxation rates can be different. The
results presented here should therefore motivate the development
of a detailed theory of relaxation effects in Si:B. Independent of
this, we have performed microwave power saturation measurements
(not shown) and confirmed that all resonances in Si:B saturate at
about the same microwave power. The multiple relaxation pathways
in the four level system make a direct interpretation of these
saturation experiments difficult. Nevertheless, the fact that
microwave power saturation occurs at the same power indicates that
at least some of the $\gamma_{ij}$ in Si:B have comparable magnitudes
as assumed.

Our model is still somewhat qualitative and does not reproduce all
experimental facts. For example, the model predicts that the
strongest effect from double excitation should be seen at close to
saturation conditions, i.e. $\gamma_{ij}/w$$ \sim$1, which is not
in line with experiment. Further improvements to the model could
be based on the density matrix formalism and the Redfield
relaxation theory to construct the full set of Bloch equations for
the $J$=3/2 spin system, also taking into account the specific
spin Hamiltonian of Si:B to correctly predict the relaxation rates
between different spin levels. Nevertheless, the instructive model
constructed here establishes that the substructures
in the EPR lines arise from the double excitation among the four
levels of B acceptors.
The improved understanding of the resonance properties of acceptors and the
additional degree of freedom achieved via isotope engineering
could make these valence band-derived states versatile competitors
to the more thoroughly studied donors also in applications such as
spin- or orbitronics.

\begin{acknowledgments}
The work at Keio was supported in part by the JST-DFG Strategic
Cooperative Program on Nanoelectronics, in part by Grant-in-Aid
for Scientific Research \#18001002, by Special
Coordination Funds for Promoting Science and Technology, and by a Grant-in-Aid for the Global Center of Excellence at Keio
University. Work in Garching was supported by the JST-DFG program
(Br 1585/5). Work at Princeton was supported by the NSA/LPS
through LBNL (MOD 713106A).
\end{acknowledgments}


\end{document}